\RequirePackage{fix-cm}
\documentclass[smallextended]{svjour3}       
\smartqed  
\usepackage{graphicx}
\usepackage{hyperref}
\usepackage{cite}
\usepackage{amsmath}
\usepackage{algpseudocode}
\usepackage{algorithm}
\DeclareMathOperator*{\argmax}{arg\,max}

\begin{document}

\title{On community structure in complex networks: challenges and opportunities
}

\titlerunning{Community structure: challenges and opportunities}        

\author{Hocine Cherifi         \and
        Gergely Palla       \and 
        Boleslaw K. Szymanski  \and 
        Xiaoyan Lu
}

\authorrunning{Cherifi, Palla, Szymanski, Lu} 

\institute{ Hocine Cherifi \at
              LIB EA 7534 University of Burgundy, Esplanade Erasme, Dijon, France \\
              \email{hocine.cherifi@u-bourgogne.fr}           
           \and
           Gergely Palla  \at
              MTA-ELTE Statistical and Biological Physics Research Group \\
              P{\'a}zm{\'a}ny P. stny. 1/A, Budapest, H-1117, Hungary\\
              \email{pallag@hal.elte.hu}
        \and
           Boleslaw K. Szymanski \at
              Department of Computer Science \& Network Science and Technology Center\\
              Rensselaer Polytechnic Institute\\
              110 $8^{th}$ Street, Troy, NY 12180, USA\\
              \email{szymab@rpi.edu}
            \and 
            Xiaoyan Lu \at
              Department of Computer Science \& Network Science and Technology Center\\
              Rensselaer Polytechnic Institute\\
              110 $8^{th}$ Street, Troy, NY 12180, USA
            \email{luxy622@gmail.com}
}

\date{Received: November 5, 2019}

\maketitle

\begin{abstract}
  Community structure is one of the most relevant features encountered in numerous real-world applications of networked systems. Despite the tremendous effort of a large interdisciplinary community of scientists working on this subject over the past few decades to characterize, model, and analyze communities, more  investigations  are  needed  in  order  to  better  understand  the  impact  of community structure and its dynamics on networked systems. Here, we first focus on generative models of communities in complex networks and their role in developing strong foundation for community detection algorithms. We discuss modularity and the use of modularity maximization as the basis for community detection. Then, we follow with an overview of the Stochastic Block Model and its different variants as well as inference of community structures from such models. Next, we focus on time evolving networks, where existing nodes and links can disappear, and in parallel new nodes and links may be introduced. The extraction of communities under such circumstances poses an interesting and non-trivial problem that has gained considerable interest over the last decade. We briefly discuss considerable advances made in this field recently. Finally, we focus on immunization strategies essential for targeting the influential spreaders of epidemics in modular networks. Their main goal is to select and immunize a small proportion of individuals from the whole network to control the diffusion process. Various strategies have emerged over the years suggesting different ways to immunize nodes in networks with overlapping and non-overlapping community structure. We first discuss stochastic strategies that require little or no information about the network topology at the expense of their performance. Then, we introduce deterministic strategies that have proven to be very efficient in controlling the epidemic outbreaks, but require complete knowledge of the network.
\end{abstract}

\section{Introduction}

Complex systems are found to be naturally partitioned into multiple modules or communities. In the network representation, these modules are usually described as groups of densely connected nodes with sparse connections to the nodes of other groups. When a node can belong to a single community the community structure is said to be non-overlapping, while in overlapping communities a node can belong to multiple communities.
In this position paper, in three subsequent sections, we discuss three fundamental questions tied to the community structure of networks: generative models, communities in time evolving networks and immunization techniques in networks with modular structure.

In the next section, we review the work on generative models for communities in complex networks and their role in developing strong foundation for community detection algorithms. We start with modularity which is an elegant and general metric for community quality, and which has also been used as the basis for community detection algorithms by modularity maximization~\cite{newman2006Modularity, Clauset2004Finding, Blondel2008Fast, Chen2014Community}. This method was recently proven~\cite{newman2016equivalence} to be equivalent to maximum likelihood methods for the planted partition. More generally, the recovery of stochastic block model finds the latent partition of networks nodes into the communities which are equal to or correlate with the truth communities used for generation of the given network.

The stochastic block model also serves as an important tool for the evaluation of community detection results, including the diagnosis of the resolution limit on community sizes and determining the number of communities in a network. We review several widely used random graph models and introduce the definitions of the stochastic block model and its variants. We also described some recent results in this area. The first one presented in~\cite{lu2019asymptotic} discovers sufficient and necessary conditions for modularity maximization to suffer from resolution limit effects and proposes a new algorithm designed to avoid those conditions. Another one, presented in~\cite{lu2019Regularized}, uses one parameter to indicate if the assortative or disassortative structure is sought by the inference algorithm. This approach enables the algorithm to avoid being trapped at the inferior local optimal partitions, improving both computation time and the quality of the recovered community structure.   

Section \ref{sec:Time} focuses on the time evolution of complex systems, study of which has been enabled by the rapid increase in the amount of publicly available data, including time stamped and/or time dependent data. The network representation of such systems naturally corresponds to time evolving networks, where existing nodes and links can disappear, and in parallel new nodes and links may be introduced. The extraction of communities under such circumstances poses an interesting and non-trivial problem that has gained considerable interest over the last decade. Over time, communities might grow or shrink in size, may split into smaller communities or merge together forming larger ones, absolutely new communities may also emerge, and old ones can disappear. Keeping track of a rapidly changing community embedded in a noisy network can be challenging, especially when the time resolution of the available data is low. Nevertheless, considerable advances have been made in this field over the years, which we shall briefly discuss.

Section \ref{sec:Immunization} focuses on immunization strategies designed for modular networks. It  is motivated by the importance of prevention of epidemic whose outbreaks, such as diseases, represent a serious threat to human lives and could have a dramatic impact on the society~\cite{mat1, mat2}. Immunization through vaccination permits to protect individuals and prevent the propagation of contamination to their neighbors. As mass vaccination is not possible when there is limited dose of vaccines designing efficient immunization strategy is a crucial issue. Immunization strategies are  the essential techniques to target the influential spreaders in networks. Their main goal is to select and immunize a small proportion of individuals from the whole network to control the spread of epidemics. To do so they rely on various properties of the network topology. For example, the network degree distribution has been extensively studied. Indeed, as real-world contact networks exhibit a power-law degree distribution, targeting preferentially high degree nodes appears to be an effective strategy. Community structure is also a well-known property of social networks.  Recent studies have shown that it affects the dynamics of epidemics, and that it needs to be considered to design tailored epidemic control strategies~\cite{cc,cbf,bhd,ress,commbet}. This section presents an overview of recent and influential works on this issue.


\section{The generative models of communities in complex networks}

\subsection{Model definition} \label{sec:models}

\subsubsection{Erd{\H o}s--R{\'e}nyi model}
The Erd{\H o}s--R{\'e}nyi (ER) random graph~\cite{erdds1959random} is perhaps one of the earliest works on random graph models. It has two closely related definitions. Given a set of $n$ nodes and $m$ edges, one variant of the ER model randomly connects $m$ pairs of different nodes. This process generates a collection of unique graphs of exactly $n$ nodes and $m$ edges, each of them being generated uniformly at random.

The other variant of ER model~\cite{gilbert1959random} specifies the probability of forming an edge between every pair of different nodes. According to this definition, each pair of nodes is connected with a probability $p$ independently at random. By the law of large numbers, as the number of nodes in such random graph tends to infinity, the number of generated edges approaches $\binom{n}{2} p$. The likelihood of generating a network $G$ of $n$ nodes and $m$ edges is
\begin{equation}
    P[G] = p^m \left(1- p\right)^{\binom{n}{2} - m}
\end{equation}
Since every edge is generated randomly with the same probability $p$, the degree of any particular node in the ER model follows the Binomial distribution.

\subsubsection{Configuration model}
Similar to the ER model, the configuration model~\cite{molloy1995critical} assumes that the edges are placed randomly between the nodes. The randomization conducted by the configuration model always preserves the pre-defined node degree which can be represented as the number of adjacent half-links or stubs. The network generation process keep randomly pairing every two stubs to create an edge until no stub remains. Hence, the configuration model produces an ensemble of graphs with the same degree sequence. The number of edges between nodes $i$ and $j$ averaged over all the graphs generated in this way is equal to $\frac{k_i k_j}{2m}$ where $k_l$ is the degree of node $l$ and the number of edges $m = \frac{1}{2} \sum_l k_l$. The configuration model is considered a benchmark in the calculation of modularity~\cite{newman2016equivalence}, a commonly used quality metric for network partitions. Given a partition of network nodes into communities, modularity compares the number of edges observed in each community with the corresponding expected number in the graphs generated by the configuration model with the same degree sequence, which is given as
\begin{equation} Q = \frac{1}{2m} \sum_{r} \sum_{\{i,j\} \in r} \left(A_{ij} - \frac{ k_i k_j }{ 2m } \right) = \sum_{r} \left[\frac{m_r}{m} - \left(\frac{ \kappa_r }{ 2m } \right)^2 \right],
\end{equation}
where $\{i,j\} \in r$ denotes every pair of nodes inside community $r$, $m_r$ is the number of edges with both endpoints inside the community $r$, $\kappa_r$ is the sum of the degrees of nodes in community $r$.
 
It is worth noting that the network generated by the configuration model does not exclude the self-loop edges, each of which connects a node to itself and the multi-links which are the multiple edges between the same pair of nodes. However, when the number of nodes approaches infinity, the density of self-loops and multi-links in the network generated by the configuration model tends to zero.

Unlike the pre-defined node degrees in the configuration model, the expected node degrees in the ER model are all the same, which are rarely observed in real graphs. Thus, the graphs produced by the configuration model are more realistic than the ER graphs thanks to the node degree variations.

\subsubsection{Stochastic block model}

\textbf{Standard SBM.} The standard stochastic block model~\cite{holland1983stochastic} is a generative model of the graph in which nodes are organized as blocks and edges are placed between nodes independently at random. In the standard stochastic block model, each node $i$ in the network is associated with a block assignment $g_i$. The number of edges between nodes $i$ and $j$ is independently distributed. It follows a Bernoulli distribution with mean $\omega_{g_i,g_j}$, a parameter which depends only on the block assignments of two endpoints. Thus, the standard stochastic block model is parameterized by a matrix $\Omega=\{\omega_{rs}\}$ whose component $\omega_{rs}$ denotes the probability of forming an edge between a node in block $r$ and the other node in block $s$. Given the block assignment $\{g_i\}$ and the edge probability matrix $\Omega$, the likelihood of generating an undirected unweighted network $G$ is
\begin{equation}
    P[G|\{g_i\}, \Omega] = \prod_{i<j} \left( \omega_{g_i g_j} \right) ^ {A_{ij}} \left( 1 - \omega_{g_i g_j} \right)^{1-A_{ij}}
\end{equation}
where $A_{ij} \in \{0, 1\}$ denotes the number of edges between nodes $i$ and $j$. The standard stochastic block model can be used in a priori block model setting, where the block assignments are pre-defined, and the objective is to estimate $\Omega$. It can also be used in the posteriori block model setting, which estimates $\Omega$ and the block assignments $\{g_i\}$ simultaneously.

In the posteriori block modeling setting, the number of blocks in SBM is usually specified in prior to applying the statistical inference such as maximum likelihood estimation (MLE) to the model. This is because the maximum likelihood generating the observed graph always increases by assuming more blocks. When every node forms one single block, the MLE obtains $\hat{\omega}_{g_i g_j}=1$ if there is one edge between nodes $i$ and $j$ and $\hat{\omega}_{g_i g_j}=0$ otherwise, leading to the maximum likelihood $P[G|\{g_i\}, \Omega]=1$. But such specification of the model does not have any practical usage. On the other hand, when there is only one block, the $\Omega$ matrix becomes a scalar value and the standard SBM reduces to the ER model. Therefore, it is also an important task to find the number of SBM blocks in a network.\\

\noindent
\textbf{Degree-corrected SBM.}
Since the standard stochastic block model considers nodes in the same block statistically indistinguishable in terms of the probability of forming edges, the degree heterogeneity is ignored. However, real-world networks typically display broad degree distributions. The lack of degree heterogeneity makes the standard stochastic block model unsuitable for applications to many realistic networks. Therefore, the degree-corrected stochastic block model~\cite{karrer2011stochastic} incorporates the degree heterogeneity, assuming that the number of edges between any pair of nodes $i$ and $j$ follows the Poisson distribution with mean $\omega_{g_i,g_j} \theta_i \theta_j$ where $\theta_l$ is a model parameter associated with each node $l$. In an unweighted undirected multi-graph, after ignoring all terms independent of the model parameters, the log-likelihood simplifies to
\begin{equation} \label{eq:ll}
    \log P[G|\{g_i\}, \{\theta_{i}\}, \Omega] = \frac{1}{2} \sum_{ij} \left[ A_{ij} \log \left( \omega_{g_i,g_j} \theta_i \theta_j \right) - \omega_{g_i,g_j} \theta_i \theta_j \right]
\end{equation}
where $A_{ij}$ is the number of edges between different nodes $i$ and $j$ for $i\neq j$; for the simplicity of the expression, the model defines $A_{ii} = 2k$ for any node $i$ with $k$ self-loop edges. Given a partition of the network, i.e., the block assignments $\{g_i\}$, the posterior maximum likelihood estimates of $\theta_i$ and $\omega_{rs}$ are
\begin{equation}
    \hat{\theta}_i = \frac{k_i}{\kappa_{g_i}}, \quad \quad \hat{\omega}_{rs} = m_{rs},
\end{equation}
where $\kappa_r = \sum_{i\in r} k_i$ is the sum of the degrees of all nodes in a block $r$, and $m_{rs}$ is the total number of edges between blocks $r$ and $s$, or twice the number of edges in $r$ if $r=s$. Plugging in the maximum likelihood estimates above and skipping the irrelevant terms, the log-likelihood of the degree-corrected stochastic block model can be simplified as
\begin{equation}
    \log P[G|\{g_i\}] = \sum_{rs} m_{rs} \log \frac{m_{rs}}{\kappa_{r}\kappa_{s}} .
\end{equation}
It is worth mentioning that the degree-corrected stochastic block model assumes that the number of edges between any two nodes follows the Poisson distribution. In the standard stochastic block model where the number of edges draws from the Bernoulli, it is rare that the edge probability is close to 1, because most real networks are often sparse. A Bernoulli random variable with a small mean is well approximated by a Poisson random variable~\cite{perry2012null}, which makes the Poisson distribution a good replacement here for the number of edges between two nodes.

\subsubsection{Planted partition model}
The standard planted partition model~\cite{mcsherry2001spectral,condon2001algorithms} is a special case of the standard stochastic block model. The network generated by the planted partition model includes an edge between any two nodes in the same block with a probability $p$ and an edge between any two nodes across different blocks with a probability $q$.
When $p > q$, the network generated by the planted partition model has an assortative structure; otherwise, when $p<q$, the model generates networks with disassortative structure, which corresponds to the bi-partite networks~\cite{asratian1998bipartite} when only two blocks exist.

Similar to the degree-correction of the standard stochastic block model, the standard planted partition model can be extended to its degree-corrected version. In the degree-corrected planted partition model~\cite{newman2016equivalence}, the number of edges between any two nodes $i$ and $j$ follows the Poisson distribution with mean $\omega_{g_i,g_j} \frac{k_i k_j}{2m}$ where $\omega_{g_i,g_j}=\omega_1$ if $g_i = g_j$ or otherwise $\omega_{g_i,g_j}=\omega_0$. Given the block assignments $\{g_i\}$ and parameters $\omega_0$ and $\omega_1$, the log-likelihood of generating a particular graph is
\begin{align} \label{eq:logl_ppm}
    \log P[G|\{g_i\}, \{\omega_1, \omega_0\}] &= \frac{1}{2} \sum_{ij} \left[ A_{ij} \log \left( \omega_{g_i,g_j} \frac{k_i k_j}{2m} \right) - \omega_{g_i,g_j}\frac{k_i k_j}{2m} \right]
\end{align}
which, after a small amount of manipulation, can be re-written as
\begin{equation} \label{eq:ppm_modularity}
     \log P[G|\{g_i\}, \{\omega_1, \omega_0\}] = B \left[ \frac{1}{2m} \sum_{r} \sum_{\{i,j\} \in r} \left(A_{ij} - \gamma \frac{ k_i k_j }{ 2m } \right) \right] + const.
\end{equation}
where $\{i,j\}\in r$ denotes every pair of nodes in block $r$, the terms $B=m\log\frac{\omega_1}{\omega_0}$ and $\gamma = \frac{\omega_{1}-\omega_{0}}{\log \omega_{1} - \log \omega_{0}}$ are independent of the block assignments $\{g_i\}$. Comparing Eq.~\ref{eq:ppm_modularity} with the definition of generalized modularity of Reichardt and Bornholdt~\cite{reichardt2006statistical}, maximizing the log-likelihood of the degree-corrected planted partition model is equivalent to maximizing the generalized modularity with a specific resolution parameter $\gamma$. This equivalence result shows that maximizing generalized modularity tends to find communities of similar statistical properties. In realistic networks where edges are heterogeneously distributed within different communities, however, there may not be a single resolution parameter $\gamma$ sufficient to avoid the resolution limit anomaly~\cite{fortunato2007resolution, lu2019asymptotic}. As a result, small well-formed communities are likely to be merged into inappropriate large groups, while large well-formed communities spread across smaller ones.

\subsection{Model Inference}
Despite of being widely used for community detection, modularity maximization is provably
NP-Hard~\cite{Brandes2006Maximizing} that implies that any algorithm based on this approach may fail on some inputs. It also suffers from the resolution limit anomaly in which the well-formed dense communities get merged into a large cluster or the loose community inappropriately splits into multiple smaller clusters to increase the modularity. An alternative approach for community detection is the statistical inference to fit the generative model to the observed network data. Such approach assumes the observed network is produced by a random graph model with a pre-defined partition of the network as the model parameter. In general, the statistical inference aims at recovering the partition which maximizes the likelihood of the random graph model generating the observed network data. In this section, we introduce the inference methods for the generative graph models which usually requires selecting the number of blocks and discuss their connection to the traditional modularity optimizations and the resolution limit anomaly in Section~\ref{sec:MLE_resolution_limit}.

\subsubsection{\bf Selecting the number of communities}
The stochastic block model and its variants do not specify the number of communities in the network. In general, the likelihood of these models increases as the number of communities grows. Thus, maximizing the likelihood of the model produces the trivial results where every node becomes a single community. Therefore, one needs to specify the number of communities for these random graph models. One approach is to find the number of communities by the statistical inference~\cite{riolo2017efficient, newman2016estimating}. Alternatively, according to the Occam's Razor, the model inference process should take into account the complexity of the model, which can be measured by the model description length~\cite{peixoto2012entropy}. Other work~\cite{peixoto2017bayesian} also uses the Bayesian model selection to determine the number of the communities in a network.~\cite{ghasemian2019evaluating} provides a detailed discussion of commonly used approaches to select the number of communities for random graph models.

\subsubsection{Monte Carlo Markov Chain}
The simplest Markov Chain Monte Carlo approach is to propose moving each node from its original block into one of the $B$ blocks randomly, which easily satisfies the requirements of ergodicity and detailed balance because any block assignment can be reached from the current block assignment with finite and aperiodic expected number of steps. However, considering the size of the partition space $O(N^K)$ for a network with $N$ nodes and $K$ blocks, the naive MCMC approach is not practical. Therefore, ~\cite{peixoto2014efficient} proposes the optimized Markov Chain Monte Carlo (MCMC) algorithm with the greedy heuristic to infer the block assignment. Initially, every node in the network is assigned to one random block independently. Then, one attempts to move a node from block $r$ to $s$ with a probability conditioned on its neighbor's block assignment $t$
$$p(r\to s|t) = \frac{m_{ts} + \epsilon}{\sum_s m_{ts} + \epsilon B}.$$
In the above, $\epsilon > 0$ is a free parameter to fulfill the ergodicity condition such that any block assignment can be reached from the current block assignment with the finite and aperiodic expected number of steps. When $\epsilon$ tends to $\infty$, the proposed function reduces to the naive scheme which assigns random block to the current node. However, such naive scheme is inefficient. Indeed, the possibility of current node being assigned to the correct block assignment is very low, thus, such assignment does not increase the log-likelihood in most cases. Consequently, the assignments are rejected very frequently, wasting the computational resource. By applying a relatively small $\epsilon$, the assignment selected by the function proposed above is more likely to get accepted. The intuition behind this function is that, given that there are many edges across blocks $s$ and $t$, a node with many neighbors in block $t$ is likely to be assigned to block $s$. Thus, the function proposed above is more likely to be accepted, avoiding the computational cost wasted by many rejected assignments.

To ensure the detailed balance, each proposed move is accepted with a probability $a$ in the Metropolis-Hastings fashion~\cite{metropolis1953equation} given by
\begin{equation}
a = \min \Bigl\{ \exp(\Delta \mathcal{L}) \frac{\sum_t n_t p(s\to r|t)}{\sum_t n_t p(r\to s|t)} \Bigr\}  ,
\end{equation}
where $\Delta \mathcal{L}$ is the change of log-likelihood after the move and the node of the proposed move has $n_t$ neighbors in block $t$.

In~\cite{lu2019Regularized}, the authors observe that the current versions of stochastic block model randomly search through the large space of potential solutions containing both assortative and disassortative structures. Consequently, inference algorithms using these models are often trapped in a solution unsuitable for the user and it takes them long time to escape. To address this issue, the authors of~\cite{lu2019Regularized} apply a simple constraint on nodes’ internal degree ratio in the objective function. This approach is independent of the inference algorithm. The resulting algorithm reliably finds assortative or disassortative structure as directed by the value of a single parameter. The paper contains the results of validation of the model experimentally by testing its performance on several real and synthetic networks. The experiments show that the inference of degree-corrected stochastic block model quickly converges to the desired assortative or disassortative structure. In contrast, the inference of degree-corrected stochastic block model gets often trapped at the inferior local optimal partitions.

\subsubsection{Modularity optimization method}
\cite{newman2016equivalence} proposes an iterative algorithm to find the optimal values of $\Omega, {\bf g}$ that maximize the log-likelihood of the degree-corrected planted partition model. The author of \cite{newman2016equivalence} shows the maximum likelihood estimates of the block assignments ${\bf g} = \{g_i\}$ is equivalent to maximizing the generalized modularity
\begin{equation}
    Q(\gamma) = \frac{1}{2m} \sum_{ij} (A_{ij} - \gamma \frac{ k_i k_j }{ 2m } ) \delta_{g_i,g_j}
\end{equation}
which is given as a function of $\gamma$, a positive parameter known as the resolution parameter. The algorithm repeats the following two steps until convergence:
\begin{itemize}
    \item Given the values of $\Omega = \{\omega_1, \omega_0\}$, find the optimal block assignment ${\bf g}$ maximizing the log-likelihood of degree-corrected planted partition model defined in Eq.~\ref{eq:logl_ppm}. This is equivalent to maximizing the generalized modularity $Q(\gamma)$ with a $\gamma = \frac{\omega_{1}
-\omega_{0}}{\log \omega_{1} - \log \omega_{0}}$,
    $${\bf g}^{\text{new}} = \argmax_g \log P({\bf A}|\Omega, {\bf g}) = \argmax_g Q(\gamma) $$
    \item After updating ${\bf g}$, find the $\Omega=\{\omega_{\text{1}}, \omega_{\text{0}}\}$ under the current block assignment ${\bf g}$ by the maximum likelihood estimation, $$\Omega^{\text{new}} = \argmax_\Omega \log P({\bf A}|\Omega, {\bf g}) $$
\end{itemize}

\subsubsection{Relation to modularity resolution limit} \label{sec:MLE_resolution_limit}
The maximization of the generalized modularity is equivalent to the maximum-likelihood estimation (MLE) of the degree-corrected planted partition model on the same graph~\cite{newman2016equivalence}. Hence, the partition of the network which most likely generates the observed network also maximizes the generalized modularity with a particular resolution parameter. However, in the planted partition model, all communities have similar statistical properties, which is unusual in practical application. 

In~\cite{lu2019asymptotic}, the authors answer the important question about the performance of the generalized modularity on the networks generated by the stochastic block model that can generate more realistic networks with heterogeneous communities.
First, these authors establish an asymptotic theoretical upper and lower bounds on the resolution parameter of generalized modularity bridging the gap between the literature on the resolutions limits of modularity-based community detection~\cite{fortunato2007resolution} and the random graph models. They also show that communities with different densities can still be detected by maximizing the generalized modularity when the resolution parameter is within the established range. Otherwise, when this parameter is larger than the upper bound established in this paper, some well-formed communities are likely to be spread among multiple clusters. In the opposite case when the resolution parameter is lower  than the bound presented in the paper, some communities are inappropriately merged into one large component.
The conclusion is that the generalized modularity resolution limits arise when a network contains a subgraph whose lower bound is higher than the upper bound of another subgraph because in such a case any resolution parameter will be either above the upper bound of latter subgraph or below the lower bound of the former subgraph or both.

To address the above mentioned problem, the authors of~\cite{lu2019asymptotic} introduce a progressive agglomerative heuristic algorithm that systematically increases the resolution parameter. The algorithm recursively splits the resulting clusters of the previous level to detect smaller communities. As the recursion proceeds, the algorithm gradually increases the resolution parameter for high-resolution community detection in local subgraphs of the network. The algorithm proceeds until the final partition is no longer statistically significant. This approach avoids getting trapped by the resolution limit and does not require multiple re-computing of the resolution parameter~\cite{newman2016equivalence}, which can be computationally prohibitively costly for large networks.

\section{Time evolving communities}
\label{sec:Time}

As mentioned in the Introduction, one of the challenging problems related to communities is given by the partitioning of time evolving networks. Here we briefly overview the most widely used methodologies and important advances related to this area. A very nice survey providing a more in depth description of the various approaches with formal definitions, algorithms, etc. was recently published by Rossetti and Cazabet in Ref.\cite{Rossetti_survey}.

\subsection{Snapshot based approaches}

Probably the most simple approach is to define snapshots, corresponding to static graphs, representing the state of the evolving network at a given time point, and to apply a static community finding method to the subsequent snapshots \cite{Hopcroft_evolv_coms,Asur_evolv_coms,Palla_com_evolv,Greene2010TrackingTE,Martin_evolv_coms,Brodka_GED}. The communities found in the neighboring time steps then have to be matched with each other somehow. One of the basic ideas is to use the Jaccard-index for measuring the relative overlap between the communities, and match the pairs in the decreasing order of the Jaccard-index \cite{Palla_com_evolv}. Naturally, the Jaccard-index can be replaced by any other similarity measure such as e.g., the normalized mutual information \cite{Danon_mutinfo,Lancichinetti_mutinfo}, the adjusted mutual information \cite{Amelio_adjusted_mutinfo}, or any advanced information based similarity in general. 

The advantage of this approach is that it is conceptually simple, and one can use basically any community finding method  on the static snapshots. The drawback is that the matching part can become technically complicated under certain circumstances. First of all, if there are ${\cal O}(N_c)$ communities found in a given snapshot, in principle we need to evaluate the chosen similarity function ${\cal O}(N_c^2)$ times for every pair of subsequent snapshots. Moreover, for similarity measures based on solely memberships (without taking into account e.g., the link structure of the communities) it is not uncommon for a community $C_i(t)$ at time step $t$ to have two or even more corresponding communities $C_j(t+1)$ at time step $t+1$ with equal similarity to $C_i(t)$ simply because the membership values can take only integer numbers. Thus, when choosing the most similar community from the next time step as the image of $C_i(t)$ at $t+1$, we might run into the problem of having multiple equally similar candidates. Another problem is that a large community at $t$ can have a non-zero similarity with many different communities at $t+1$, and thus, if we follow the merging and splitting processes between the communities without any restriction on the minimal similarity, the lineage of the evolving community structure can become extremely subtle and complicated. Of course, using a minimum similarity threshold can make the picture clearer, however, at the cost of the introduction of an extra parameter to the method. Last but not least, in case we are using a static community finding method that allows overlaps between the communities, finding the best match between the subsequent time steps can become even more tricky \cite{Palla_com_evolv}. For the above reasons, the introduction of more specialized community finding methods targeted at time dependent networks was very well motivated. 

\subsection{Evolutionary algorithms}
The key idea behind these approaches is to provide a unified framework in which the inference of communities at a given time step $t$ can take into account information about the network structure at other time steps as well. One of the first methods pointing in this direction was suggested in \cite{Chakrabarti_evolutionary_clustering}, where the goal was to optimize both for 'point wise' precise communities reflecting the modular structure of the network at any given time point $t$, and for keeping the change in the community structure between two subsequent time steps as low as possible. This was achieved in a rather general framework, where a cost function is introduced composed of two parts, the first related to the accuracy of the communities located at the different time steps, and the second term corresponding to the 'historical cost', depending on the similarity of the partitions at subsequent time steps. The second term also involves a user defined parameter (a simple multiplicative factor) with which we can balance the trade-off between lowering the point-wise accuracy and gaining smoothness of evolution in time. In \cite{Chakrabarti_evolutionary_clustering} the method is used with hierarchical clustering and $k$-means clustering together with historical costs specifically using the nature of the applied clustering. 

In principle, the above framework can be used with any static community finding algorithm combined with a suitable similarity measure between communities. E.g., in \cite{Chi_evolv_spectral_clust} spectral clustering techniques are used to uncover the communities, whereas in Ref.\cite{Lin_evolv_coms}, the community finding is based on optimizing the Kullback--Leibler divergence between the actual network structure and the one predicted based on community memberships. The advantage of this latter approach is that the historical costs can also be formulated as the Kullback--Leibler divergence between the consecutive community partitions, providing a unified formulation for both type of costs, and also allowing for a probabilistic interpretation of the optimization problem \cite{Lin_evolv_coms}. Further methods similar in nature were proposed in Refs.\cite{Zhou_evolv_coms,Tang_evolv_coms,Folino_evolv_coms,Sun_evolv_coms,Gong_evolv_coms,Kawadia_evolv_coms,Crane_evolv_coms,Gorke_dyn_coms_smooth}.

Another quite general framework for evolutionary community finding was proposed in \cite{Mucha_multiplex_coms}, based on the concept of multislice networks. In such systems, the network structure can be organized into layers, where the layers represent different types of connections between the same nodes such as e.g., social media connections, e-mail connections and personal acquaintances between the same people. By taking any community finding approach in general that is suitable for detecting communities in multiple levels simultaneously, the same method can be also applied to evolutionary community finding if we represent the time evolving network as a multislice network, where the different layers correspond to the subsequent time steps during the time evolution. The solution offered in \cite{Mucha_multiplex_coms} is based on modularity, however as mentioned above, the generality of the framework allows any further multislice methods as well.

A further general problem class into which the challenge of evolutionary clustering fits naturally is given by consensus clustering \cite{Lancichinetti_consensus_clust}. The basic idea of consensus clustering is to apply multiple different clustering methods to the same network, and then bring the found (presumably different) partitions to consensus, resulting in stable, relevant communities even for stochastic community finding methods. However, this approach is also very suitable for evolutionary clustering when the setup is modified as follows. First, based on the time evolving network data, following the well-known concept of sliding time windows, a number of time frames are defined, where each frame corresponds to the aggregation of a certain number of consecutive time steps in the original data, and also the neighboring time frames show a significant overlap with each other to ensure stability and a smooth time evolution of the communities. Next, a static community finding algorithm is applied to the subsequent time frames, and then the obtained results are brought to consensus, again, over sliding windows of a fixed length \cite{Lancichinetti_consensus_clust}. 

Generative models such as the stochastic block model can also provide  very interesting solutions for evolutionary clustering. In Ref.\cite{Yang_dynamic_SBM} the concept of the dynamic stochastic block model is introduced, where in addition to the usual group membership probabilities and membership dependent linking probabilities, further probabilistic transition matrices are considered for describing the evolution of node memberships between the subsequent time steps. A more general formulation of the model is given in \cite{Tiago_time_varying} with the help of a layered stochastic block model, where the layers can naturally correspond to time steps in case of a dynamic network, however the approach can handle general multilayer networks as well. Important results on the detectability thresholds for the dynamic stochastic block model are presented in \cite{Peel_detectability} based on the cavity method, while in \cite{Tiago_and_Martin_dynamic_coms}, the concept of higher order Markov chains (and thus, the possibility for memory effects) are successfully incorporated into the framework of dynamic stochastic block models. A common feature of the above methods is that the results are obtained via Bayesian inference, which in practice is usually implemented with the help of a Markov chain Monte Carlo algorithms \cite{Yang_dynamic_SBM,Tiago_time_varying,Tiago_and_Martin_dynamic_coms}.

Stochastic block models can be also successful in the analysis of systems where the network structure itself should also be generated from time dependent (and possibly noisy) signals. In \cite{Leto_unobserved_detection}, an end-to-end community detection algorithm is proposed, avoiding the extraction a sequence of point estimates for the links, and providing an inference of the stochastic blocks directly from the raw data. In parallel, the stochastic block model framework can be also used for a joint reconstruction of the network structure and the communities from time varying functional data \cite{Tiago_dynamic_reconstruction}, where synergistic effects were reported, as the inferred blocks improved the reconstruction accuracy of the links, which in turn also made accuracy of the inferred communities better.

\subsection{Incremental clustering, online community finding and predicting community evolution}

In case of the previously mentioned methods,  we assumed a 'complete knowledge' about
the time evolution of the system at least on the level of the input data, thus, when inferring the communities at a given time step, information about the network structure coming from later time steps was also available, and could be made use of. A somewhat more restrictive setup is where at a given time point only the data corresponding to previous time steps can be used. Such scenario could be when small but fast changes occur in a large network, and our aim is to always give the currently best partitioning of the network into communities, which however is also likely to be quite similar to the partitioning in the previous time steps. The concept of incremental clustering fits to this setup in a natural manner \cite{Aynaud_increment_clust}, where instead of running the community finding method of our choice 'from scratch' on the current snap shot of the studied network, we consider the changes in the network structure and update the communities from the previous time step. A method following this approach was proposed in \cite{Ning_increment_clust} based on spectral clustering, while in \cite{Bansal_increment_coms,Gorke_increment_clust} modularity optimization techniques were used for a similar purpose. However, further static community finding methods such as the label propagation approach can also be adapted to this framework as shown in \cite{Xie_and_Bolek_increment_clust}, and the problem of overlapping communities can also be handled \cite{Cazabet_increment_clust}. Additional incremental clustering techniques can be found in Refs.\cite{Duan_increment_clust,Falkowski_increment_clust,Nguyen_overlap_increment_clust,Cazabet_and_Amblard_increment_clust,Gorke_incr_clust_2,Ma_increment_clust,Lee_increment_clust,Zakrzewska_increment_coms}.

An idea closely related to incremental clustering is given by the concept of online clustering in dynamical networks \cite{Aggarwal_online_clustering}. This framework considers large networks updated in a stream fashion, where changes in the communities are detected online, separated from offline community detection and exploratory querying. A somewhat different strategy for online community finding is proposed in \cite{Zanghi_online_clustering} based on expectation-maximization and the stochastic block model, and further methods are proposed in Refs\cite{Rossetti_online_coms,Tan_online_coms}. 

A closely related problem to the above described  'instantaneous' community detection methods is given by the challenge of predicting the future changes in communities for time evolving systems. The first results in this direction were related to the prediction of whether a community will grow and/or survive, or instead will disappear \cite{Leskovec_predict,Patil_predict}. In \cite{Goldberg_predict} also the predicted life span and the connection between the life span and structural properties of the communities were studied. Beside the 'ultimate fate' and life span, predicting the occurrence of change events for communities is also a relevant problem, where the usage of machine learning techniques is a natural idea. The basic idea is to build classifiers that can predict certain type of events based on various community features \cite{Brodka_predict,Gliwa_predict,Takaffoli_predict}. A detailed study of the problem together with a thorough testing of methods on multiple real datasets is presented in \cite{Saganowski_predict}.

\section{Immunization strategies}
\label{sec:Immunization}

Various strategies have emerged over the years suggesting different ways to immunize nodes \cite{centralities}. Yet, finding even more highly effective strategies must be pursued since any improvement can play a major role in saving human lives and resources. Immunizing nodes at random is the simplest approach. This strategy has proven to be impractical since it requires a large proportion of nodes to be immunized to mitigate the epidemic spreading. To solve this problem, researchers try to come up with the best possible way to immunize a small number of key nodes using various topological features of networks. 
Up to now, these immunization strategies fall into two categories: stochastic and deterministic. In stochastic strategies, targeted nodes are identified by collecting information locally from randomly selected nodes in the network. They are totally agnostic about the full network structure. The most popular strategy in this category is the so-called Acquaintance immunization. It aims to vaccinate nodes which are randomly picked several times among the neighbors of randomly selected nodes. There is obviously a high chance that nodes with high degree are selected by the acquaintance strategy. Deterministic strategies, on the other hand, assume the knowledge of the whole network. These strategies determine the succession in which nodes of a network should be immunized to mitigate the epidemic spreading. They rank all nodes according to a given centrality measure. From high to low, nodes are targeted based on their rank.  
Deterministic strategies have proven to be very efficient in controlling the epidemic outbreaks. Their only drawback is their high requirement of the global topology of the network. This makes them impractical in large scale networks. Stochastic strategies, however, have the advantage of requiring only little information of the network at the expense of their performance, which is lower as compared to the deterministic immunization.
The standard centrality measures designed for complex networks with non-modular structure highlight different characteristics of the nodes depending upon their objective criteria. The Degree-based strategy targets highly connected nodes (hubs). The immunization of hubs results in a big reduction in network density which reduces the epidemic diffusion. It is a very efficient strategy in scale-free networks due to the power law degree distribution. The Closeness-based immunization strategy selects nodes with least average propagation length in the network as the most influential spreaders. Targeting these nodes may increase the average paths length in the network, hence the decrease of the epidemic propagation. Further, the Betweenness-based strategy immunizes nodes with maximum fraction of shortest paths passing through it. These nodes may have a considerable influence in networks in terms of controlling the information flow. Therefore, immunizing these nodes can stop the diffusion between many vertices due to their bridging role in the largest number of paths.
Despite the efficiency of these popular immunization strategies (Degree, Closeness and Betweenness-based strategies) on targeting influential nodes, they exhibit some limitations when applied to networks with community structure. According to recent research, community structure strongly affects the epidemic spreading process. Thus, the design of immunization strategies needs to take into consideration the community structure. Stochastic as well as deterministic strategies using information of the community structure have been proposed.  They can be classified into two groups according to the community structure model they use. The first group of strategies uses the non-overlapping community structure features. The second group is based on the overlapping community structure properties. The most widely known stochastic strategies together with deterministic strategies using advantageously the community structure are recalled in this section.

\subsection{Stochastic strategies}

\begin{figure}[h!]
\begin{center}
\includegraphics[width=13cm,height=5cm]{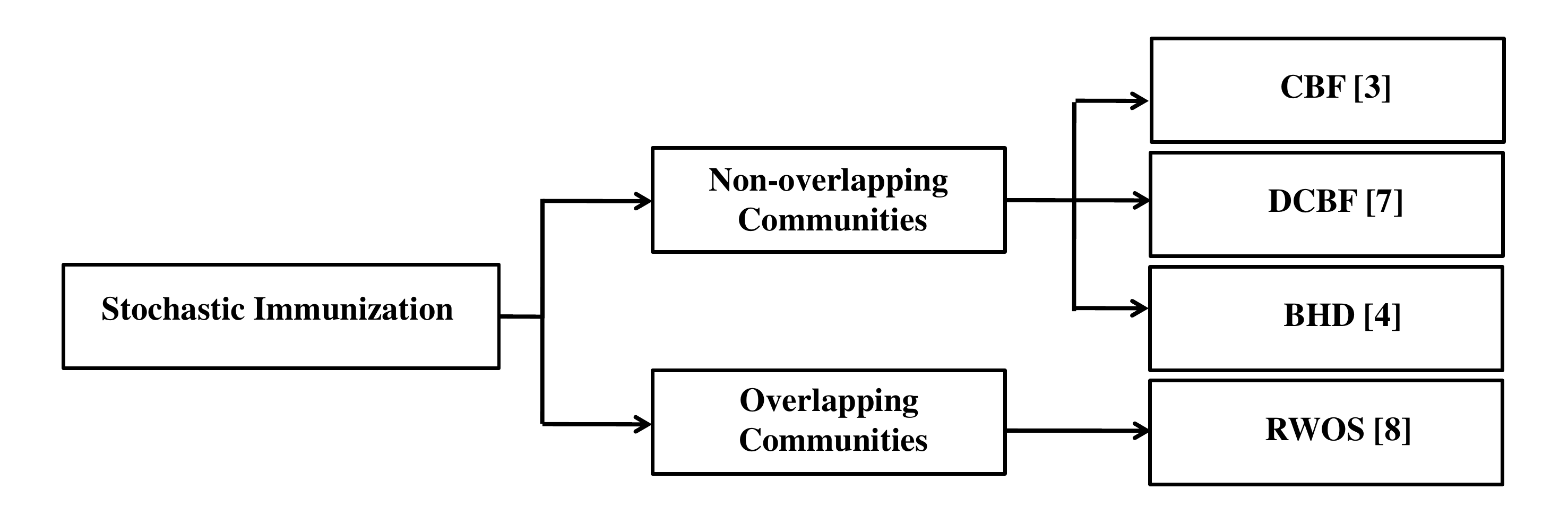}
\caption{Stochastic immunization methods.}
\label{f1}
\end{center}
\end{figure}

Stochastic immunization strategies focus on using information at the node level. They identify target nodes without knowledge of the full network structure. That makes them computationally more efficient and more practical in large networks as compared to the deterministic strategies. Roughly speaking, these strategies target either the nodes linking the communities (Bridges) or the highly connected nodes (Hubs) or the overlapping nodes using little or no information about the network topology. 

Some researchers assume that bridges are the most influential spreaders. These nodes can propagate the epidemic to the entire network because of their connectivity with various modules. They have then a global influence on the whole network and their immunization can prevent the effective diffusion to the different parts of the network. The Community Bridge Finder \textit{CBF} \cite{cbf} is an immunization strategy aiming to target the bridge nodes. It is based upon a random-walk algorithm. The community hubs are also believed to have a strong local influence in their communities. Based on this assumption the Degree Community Bridge Finder \textit{DCBF} \cite{dcbf} and the  Bridge-Hub Detector \textit{BHD} \cite{bhd} are two immunization strategies, which targets bridge nodes with high connections for immunization. The selected bridge nodes in this case play also the role of hubs. The former strategy is a variation of the \textit{CBF}, while the latter one is based on expanding friendship circles during a random walk. Other researchers try to highlight the importance of overlapping nodes in terms of the epidemic spreading dynamics. Random-Walk Overlap Selection \textit{RWOS} strategy \cite{rwos} is proposed to select the overlapping nodes according to a random-based algorithm. These key nodes can play a major role in epidemic diffusion due to their membership to multiple communities. In the following, we present a brief overview of the three stochastic strategies based on non-overlapping community structure and the one tailored for networks with overlapping community structure (refer to \autoref{f1}).


\subsubsection{Algorithms} 
 
{\bf Community Bridge Finder (CBF)}

\noindent
Immunization interventions of highly connected individuals are not always enough to protect networks from large-scale epidemics. Indeed, targeting individuals bridging communities is sometimes more effective than simply immunizing nodes with high degrees. The goal of the CBF strategy \cite{cbf} is to identify nodes acting as bridges between communities. This strategy is based on random walks. It works as follows:
 
Step 1. Select a random node $v_{i=0}$.

Step 2. Follow a random walk with the condition that a node has not been visited by the random path before.

Step 3. At each node $v_{i>=2}$, check if it is connected to more than one visited nodes. If there is just one connection, $v_{i-1}$ is considered as a potential bridge.

Step 4. Select two random neighboring nodes of $v_{i}$ other than $v_{i-1}$. If both nodes have no connections to the previously visited nodes, the node $v_{i-1}$ is then marked as a bridge and it is immunized. Otherwise, a random walk is taken back at $v_{i-1}$.

This strategy has been compared to the Acquaintance strategy defined as follows. At each step, a node is picked at random and one of its acquaintances is randomly selected, then nodes which are picked as acquaintances $n$ times are immunized. Extensive experiments were conducted on synthetic and real-world networks using SIR epidemic model. Results show that CBF outperforms mostly the Acquaintance strategy. Its best performance is obtained in networks with strong community structure (few inter community links). 

{\bf Degree Community Bridge Finder (DCBF)}

\noindent
DCBF \cite{dcbf} is a variant of the CBF strategy. The goal of this strategy is to target bridges with large amount of connections. This strategy incorporates the same steps as described in the CBF algorithm. The difference is that nodes are not randomly chosen among all the possible nodes during the random walk, but according to their degree from high to low. Two additional checks are also implemented in DCBF to decrease the computation time of the algorithm. First, the number of nodes visited in a running path is kept at the length of ten. Also, the number of visits by all random paths is recorded for each node. The node is immunized when the number of visits $k$ is equal to a certain number ($k = 2$). DCBF has been  tested on synthetic networks with various modularity values. After running the SIR epidemic model simulations, results demonstrate that DCBF performs better than the CBF algorithm in controlling outbreaks. Its performance gets higher in networks with strong community structure (when the modularity is very high $Q > 0.84$). Indeed, outbreaks are restricted locally inside communities in this case. DCBF could then target highly connected nodes in local communities, while CBF is able to identify only the bridge nodes.

{\bf Bridge-Hub Detector (BHD)}

\noindent
Communities are characterized by the heterogeneity in the connections among nodes bridging various communities. Based on this idea, BHD \cite{bhd} aims to identify bridge hub nodes as targets for immunization.  It is based on expanding friendship circles of visited nodes and works as follows

Step 1. Select a random node $v_{i=0}$.

Step 2. Follow a random walk with the condition that a node has not been visited by the random path before.

Step 3. Let $v_{i>=2}$ be the node visited after $i$ steps, and $f_{i}$ be the set of its neighbors. The node $v_{i}$ is considered as an immunization target if there is at least one node that does not take part of the set $F_{i-1}$ and that it not linked to any node in $F_{i-1}$, where $F_{i-1} = f_{0} \bigcup f_{1} \bigcup f_{2} \bigcup. . .\bigcup f_{t-1}$. Otherwise, the random walk moves on from $v_{i}$, and the friendship circle will be updated to $F_{i} = F_{i-1} \bigcup f_{i}$.

Step 4. Among the nodes in $f_{i}$, one node $v_{H}$ is randomly picked for immunization that do not belong and cannot be connected back to $F_{i-1}$.

At the end of this procedure, a pair of nodes, a bridge and a bridge hub nodes are selected for immunization. This is via friendship circles of randomly visited nodes. BHD was tested on simulated and empirical data constructed from Facebook network of five US universities using the SIR model. It results in a smaller epidemic size as compared to the Acquaintance and CBF strategies. In terms of computational time, Acquaintance is the fastest algorithm, followed by CBF and BHD. 

{\bf Random-Walk Overlap Selection (RWOS)}

\noindent
Overlapping nodes do not necessarily have high centrality measures,  yet, they can have a major effect in spreading epidemics from one community to another. Indeed, these nodes have access to multiple communities in the network. The RWOS strategy \cite{rwos} is designed to target the overlapping nodes for immunization according to a random walk. It can be specified as follows:

Step 1. Define the list of overlapping nodes.

Step 2. Select randomly a node of the network and run a random walk.

Step 3. Each visited node is nominated as a target for immunization if it belongs to the overlapping set of nodes. This process continues until reaching the desired immunization coverage.

This strategy targets highly connected overlapping nodes for immunization. It is based on the idea that the probability of visiting any node in a random path is proportional to the node degree. RWOS has been investigated on synthetic and real-world networks. The standard SIR epidemic model was run on these networks. Results show that RWOS outperforms CBF and BHD strategies in terms of the epidemic size. It performs sometimes even better than membership strategy (where nodes are immunized according to the number of communities they belong to). Moreover, its performance gets better in networks with strong community structure and higher membership values. Note that it uses more information about the community structure. Indeed, one needs to know the overlapping nodes.

\subsubsection{Discussion}

The stochastic strategies have been investigated on both simulated networks  \cite{lfr,lfrgunce} with different community structure, and real-world networks. Overall, results show that stochastic strategies based on the community structure are more efficient than the standard stochastic strategies. Results show that generally \textit{BHD} and \textit{DCBF} are more efficient than the \textit{CBF} strategy. However, \textit{BHD} strategy displays the best performance among the other strategies. Moreover, the difference between their performances increases when the modularity is high, so the communities are well separated from each other. Thus, the outbreaks stay restricted in local communities. Consequently, immunizing bridges is not an effective way to control the spreading of epidemics. That explains the poor performance of \textit{CBF} in networks with strong community structure. \textit{DCBF} may at least identify relatively highly connected bridge nodes which can cause extensive spreading of epidemics. In addition,  \textit{BHD} is capable of identifying bridge nodes with high number of inter-community links. Therefore, the effectiveness of \textit{BHD} can be attributed to the better identification of the influential spreaders as compared to the \textit{CBF} and \textit{DCBF}. All these three strategies do not take into account the overlaps between communities. On the other hand, \textit{RWOS} strategy which immunizes overlapping nodes results in smaller epidemic size as compared to the other stochastic strategies in all the networks. Furthermore, its performance enhances while increasing the membership degree of overlapping nodes. Thus, overlapping nodes play a major role in spreading infection from one community to another even if they are not necessarily highly connected.


\subsection{Deterministic strategies}
Deterministic strategies target nodes by ranking them following a given centrality measure. The centrality of a node reflects its ability to propagate the disease. The procedure of deterministic strategies can be specified as follows

Step 1. Select a given centrality measure.

Step 2. Compute the centrality for each node of the network.

Step 3. Rank nodes in decreasing order from the most to the less central node.

\begin{figure*}
\begin{center}
\includegraphics[width=15cm,height=23cm]{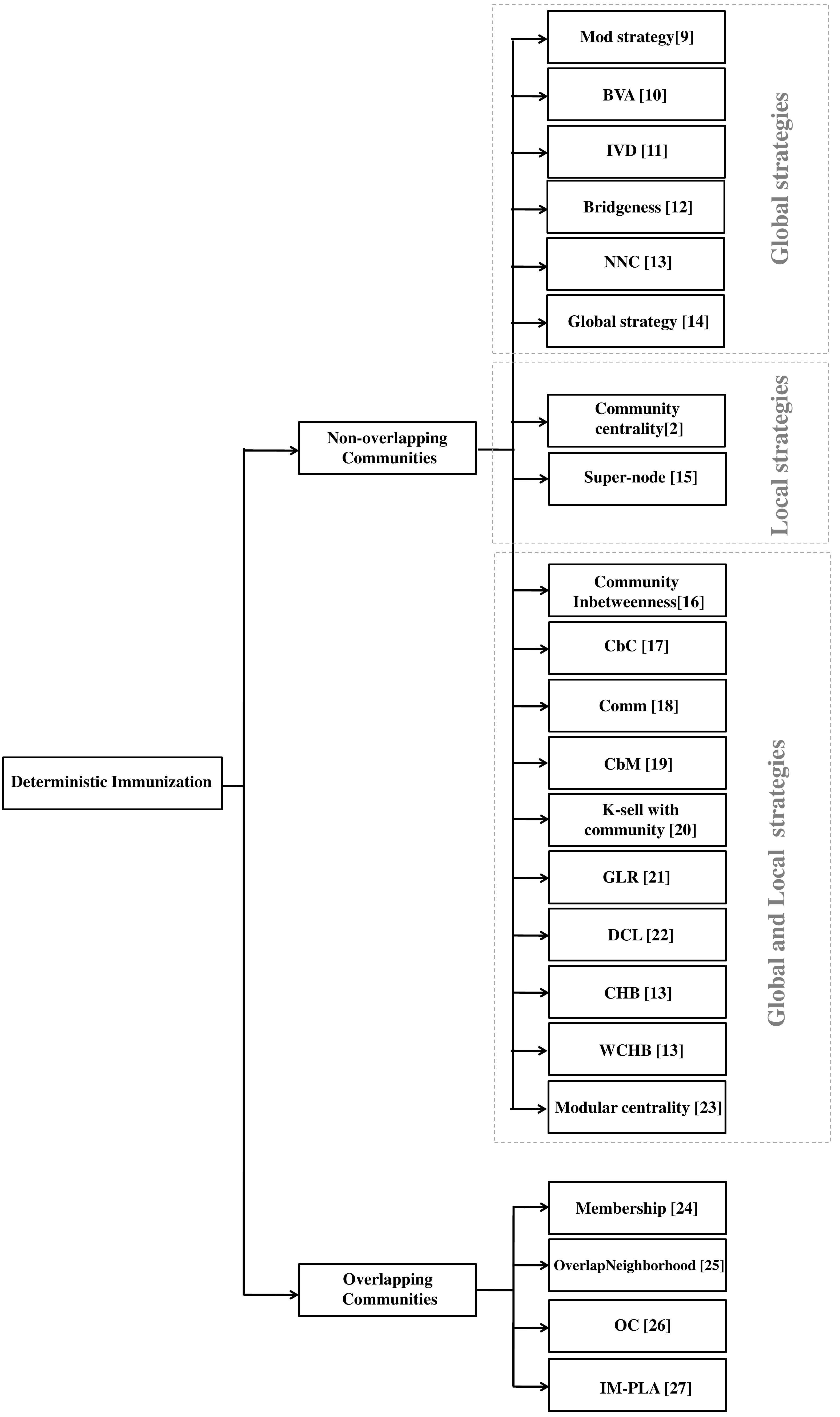}
\caption{Deterministic immunization methods.}
\label{f}
\end{center}
\end{figure*}

Step 4. Target a proportion of nodes with high ranks for immunization.\\
These strategies require the knowledge of the whole network because all the nodes are involved in the process. We now give an overview of some recent deterministic strategies designed for modular networks. They are classified into different categories according to their immunization goals (refer to \autoref{f}). 
 
\subsection{Non-overlapping community structure}
A plethora of deterministic immunization strategies are developed to identify vital nodes in networks with community structure. They can be classified into three categories (global, local,  global and local)  in networks with non-overlapping structure. The first type of strategies highlights nodes with outer connections towards foreign communities. They target bridge nodes, which can have a significant global influence on other nodes of the network. The second category tends to identify nodes with the highest local influence in their own communities. Some strategies target hubs for immunization because of their strong influence on nodes of their neighborhoods, while others immunize nodes located in the core of the community. The strategies belonging to the third category immunize both types of nodes. They select nodes having both local and global influence in the network. 

\subsubsection{Global strategies}
Bridges can be viewed as individuals that connect different subgroups of nodes in networks. They can let the epidemic outbreaks move from one module to another through their inter-community connections. Therefore, they have a major global influence in the entire network. Series of strategies have been proposed to select these critical nodes for immunization. 
The Module-based strategy (\textit{Mod strategy}) \cite{masuda} is proposed to highlight the bridge nodes between communities.  It is based on an approximated calculation of the eigenvector centrality of the coarse-grained network (called also the meta-graph). In this network the communities are represented simply by nodes, and the links are weighted by the number of links between  the two communities.  It can be specified as follows.\\

\noindent\textbf{Module-based strategy (Mod strategy):} Mod strategy was proposed by Masuda \textit{et al.} \cite{masuda}. Given the community structure of the original network, this strategy is applied on the coarse-grained network. Where each community is represented by a single node, and edges are weighted by the number of links shared by two neighboring communities. It targets nodes maximizing the following measure
\begin{equation}
Mod_{i} = 2\tilde{u}_{K} \sum_{I \neq K} d_{kI} \tilde{u}_{I}
\end{equation}
Where $\tilde{u}_{K}$ represents the eigenvector corresponding to the $K^{th}$ community. $d_{kI}$ is the number of inter-community links that exist between node $k$ and the $I^{th}$ community. The first term of this measure (i.e., $2\tilde{u}_{K}$) quantifies the importance of the community that the node $k$ belongs to, whereas the second quantity  (i.e., $\sum_{I \neq K} d_{kI} \tilde{u}_{I}$) measures its connectivity to other important communities. After immunizing all the bridge nodes, the remaining nodes are ranked according to their degree. This method preferentially targets globally important nodes having important inter-community links rather than community hubs that are locally important. The effectiveness of the Mod strategy is tested by applying it on synthetic and real-world networks of various nature. Results show that it is in most cases more efficient than Degree, Betweenness and Ress strategy (an eigenvector based strategy \cite{ress}) in networks with modular structure.\\

Different from the above method, Mantzaris \cite{bva} proposed the Boundary Vicinity Algorithm \textit{BVA}.
\noindent\textbf{Boundary Vicinity Algorithm (\textit{BVA}):} This strategy ranks nodes according to their vicinity to bridge nodes (boundary nodes) of each community. It is defined as follows

Step 1. Define the set of communities of the network.

Step 2. Extract the set of bridges which connects communities.

Step 3. Run a number of random walkers of a chosen fixed number of steps from each bridge node. Then, the number of visits to each node is counted. 

This measure quantifies the ability of a given node to propagate epidemics across bridges towards different communities. Using the SI epidemic model, the authors show that the \textit{BVA} strategy outperforms the Betweenness-based strategy in terms of the epidemic size.\\

 Yoshida \textit{et al.} proposed the Inverse Vector Density (\textit{IVD}) \cite{ivd}. It is another immunization strategy that do not require the community labels of nodes. This is by constructing a vector representation of nodes based on the modularity quality measure. The \textit{IVD} immunizes nodes with small number of nearby node vectors which are identified as bridges. This strategy performs better than the Betweenness-based strategy in terms of the Largest Connected Component (\textit{LCC}). Bridgeness strategy is proposed by Jensen \textit{et al.} \cite{bridgeness}. It is based on the Betweenness centrality while considering only shortest paths between nodes belonging to different communities. This strategy highlights nodes that connect different regions of a network. Using both synthetic and real-world networks, the Bridgeness strategy is shown to be globally more effective than the Betweenness-based strategy  to identify bridge nodes. Different from the above methods, the Number of Neighboring Community (\textit{NNC}) \cite{ourstr} selects nodes which are connected to the larger number of foreign communities, regardless of the amount of their inter-community links. It ranks nodes according to the number of neighboring communities that they can reach through at least one link. Indeed, nodes with high number of neighboring communities are able to disseminate information across the entire network. Experimental results show that the Number of Neighboring Communities strategy outperforms the Degree and the Betweenness-based strategies in terms of the epidemic size. It performs also better than some community-based strategies such as the Community Inbetweenness, \textit{CbM} strategies (see their definition in section \ref{c}). This is particularly true in networks with a community structure of medium strength (i.e., when the proportion of intra-community links is of the same order than the proportion of inter-community links). M. Kitromilidis et al. \cite{evans} propose to redefine the traditional centrality measures to characterize the influence of Western artists. This global strategy is based on computing the standard centrality measures by considering only the inter-community links of the networks. Their idea is based on the fact that influential artists usually have connections beyond their artistic movement. The Global Betweenness and Closeness strategies are compared to their classical versions. They were tested on a painter collaboration network. Experimental results show that the Global strategies allow to highlight some influential nodes who might have been missed as they do not necessary rank high in the standard measure based strategies.

\subsubsection{Local strategies}
Hubs represent the high degree nodes with the larger amount of connections that greatly exceed the average. They are a consequence of the scale-free degree distribution observed in real-world networks. In modular networks, such nodes can be found in all the communities. They have then a strong local influence on the nodes of their own communities. Newman proposed the Community centrality \cite{cc} to identify nodes that plays a central role inside communities in terms of the number of connections. 
These nodes are responsible for the maximum information flow inside their communities. He \textit{et al.} proposed the Super node strategy \cite{supernode} that can immunize nodes with the highest intra-community links (or with highest k-core index) belonging to various communities. Both strategies are described as follows:\\ 

\noindent\textbf{Community centrality (\textit{CC}):} Newman proposed a slightly different formulation of the modularity. The Community centrality \cite{cc} is derived from the eigenvectors of the modularity matrix. The modularity matrix is divided into two projections. The first dimension represents the positive eigenvectors of the modularity matrix while the second dimension represents the negative ones. Thus, the modularity can be written in terms of these vectors as follows: 
\begin{equation}
Q=\sum_{k=1}^{c}|X_{k}|^{2}-\sum_{k=1}^{c}|Y_{k}|^{2}
\end{equation}
where $c$ is the number of communities. $X$ and $Y$ are the community eigenvectors in both dimensions. The $i^{th}$ node in the community $k$ is represented by two vectors $x_{i}$ and $y_{i}$ (the $i^{th}$ rows of $X_{k}$ and $Y_{k}$ respectively).

The magnitude of a node vector $|x_{i}|$ specifies how central the node $i$ is in its community in terms of the number of connections. Thus, the node $i$ has a large positive contribution to the modularity when this measure is large. On the other hand, a higher value of $|y_{i}|$ means that the node $i$ has many connections to other nodes from foreign communities. Therefore, the Community centrality is defined to be equal to the vector magnitude $|x_{i}|$. It measures the strength with which a given node $i$ is assigned to its community.  This measure has been tested in a co-authorship network between scientists. Results show that it is not well correlated with the degree centrality. Moreover, some nodes with high Community centrality measure have relatively low degree. However, they have more connections with nodes of their communities. Thus, nodes with high Community centrality value play a central role in the spreading process in their local neighborhood.\\

\noindent\textbf{Super node strategy:} This strategy starts by ranking communities in decreasing order according to their size. After that, the node with the largest inner degree is selected from the largest community. Then, the node with the highest inner degree in the second largest community and which do not have any connections with the previous communities is selected as the second spreader. Note that there is only one previous community for the second spreader. After visiting all the communities of the network, this process is restarted again until achieving the desired number of immunized nodes. The goal of this method is to select multiple spreaders from different communities in a balanced way. SIR simulations are performed in both synthetic and real-world networks. Experimental results show that the Super node strategy results in a smaller epidemic size as compared to the Degree-based strategy. Additionally, Super node strategy proved also its efficiency while using the k-shell decomposition method in the process of finding the influential spreaders in each community.

\subsubsection{Global and local strategies}\label{c}
The immunization strategies in this category tend to target nodes that have both local and global influence. They combine the various aspects of the previous strategies to select the most influential nodes in the network. These nodes are supposed to be the main spreaders in their communities which can also disseminate the epidemics towards other modules of the network. Community Inbetweenness \cite{pietro} together with the \textit{CbC} strategy \cite{cbc} select the Hub-bridge nodes for immunization. They can be defined as follows:\\

\noindent\textbf{Community Inbetweenness strategy:} The classical betweenness needs to solve the shortest path problem of all pairs, what makes it unfeasible in large networks. Community Inbetweenness strategy \cite{pietro} is proposed to solve this problem. It is based on an entropy-based measure which approximates the betweenness centrality. It ranks nodes based solely on community information. This strategy evaluates node importance according to the proportion of its surrounding links in addition to the external links connecting it with foreign communities. The Community Inbetweenness centrality $C_{CI}$ is defined as follows:
\begin{equation}
C_{CI}(i) = k_{i} \sum_{c \in C} p_{i \rightarrow c} \; log \left( \frac{1}{p_{i \rightarrow c}} \right)
\end{equation}
Where $k_{i}$ is the degree of node $i$. $ p_{i \rightarrow c}$ is the proportion of links connecting node $i$ to the community $c \in C$. $C$ is the set of non-overlapping communities. Community Inbetweenness tends to select nodes with high connectivity and with more links to different communities. It is based on the idea that nodes with high betweenness measure are usually located between densely connected modules. These nodes are also targeted by the standard betweenness centrality. Simulation results on real-world networks show that this strategy is more efficient than the betweenness based strategy in terms of computational performance. Both strategies are also tested with the SIR model in \cite{commbet} to compare their epidemic size. Results show that Community Inbetweenness performs almost as well as the betweenness in networks with strong community structure. It is however more efficient in networks with loose community structure.\\

\noindent\textbf{Community-based Centrality (\textit{(CbC)}:} This strategy selects nodes for immunization according to their links characteristics and the size of their communities. It targets nodes that have a big impact in their communities and that can spread epidemics to nodes from other communities. It is based on a measure that evaluates the importance of node $i$ via the following formula:  
\begin{equation}
CbC_{i}= \sum_{c=1}^{m} d_{ic} \frac{S_{c}}{N}
\end{equation}
Where $d_{ic}$ is the number of links between node $i$ and other nodes in community $c$, $m$ is the number of communities in the network, $S_{c}$ is the number of nodes in community $c$, and $N$ is the size of the network.
Simulation results using the SIR model show that CbC outperforms some traditional measures such as Degree and K-shell. Moreover, CbC can also better reflect nodes importance as compared to Closeness, Betweenness and Eigenvector centralities, with much lower computational complexity.\\

Comm strategy was proposed by Gupta \textit{et al.} \cite{comm} \cite{stra5}. The aim of this strategy is to target nodes that are at the same time hubs in their communities and bridges towards other communities. It ranks nodes according to a degree-based measure. This measure is a weighted combination of the number of intra-community links and the square of the number of the inter-community links, which accounts for importance of bridge nodes. Results on synthetic and real-world networks show that the Comm strategy is more effective or at least works as well as Module-based immunization strategy, Degree and Betweenness based strategies. Community-based Mediator (\textit{CbM}) \cite{tulu} is another strategy that takes into account the internal and external density of each node. They represent the proportion of the intra and the inter-community links of a given node respectively. This strategy is based on the entropy using both densities. It uses this information to select individuals that can propagate the epidemic in their community from internal density and in other communities from external density. Experimental results demonstrate that nodes with high CbM value have a more noteworthy effect to spread epidemics in networks than nodes having a high CbC, Betweenness, Degree, PageRank or Eigenvector value. Luo \textit{et al.} \cite{kshell} proposed also the k-shell with community strategy designed for networks exhibiting a community structure. It is based on the idea that the location of a node has a big impact on the spreading process. It is a variation of the k-shell decomposition strategy, in which decomposition method is applied to the intra and the inter-community links separately. The goal is to select for immunization hubs and bridges that are located in the core of the network. Results of SIR simulations performed on Facebook network show that it outperforms the traditional k-shell decomposition, the Betweenness and Degree based strategies. Salavati et al. \cite{glr} proposed an improved version of the Closeness-based strategy designed for modular networks. It decreases also the high computational complexity of the standard closeness method. The so-called Gateway Local Rank strategy \textit{GLR} starts by ignoring the connections between communities. Then, in each community one critical node is extracted using the betweenness centrality. After that, one node with the highest inter-community links is also extracted from each community. In the last step, nodes are ranked based on the sum of their shortest paths with the extracted core and bridge nodes instead of computing their shortest paths using all the nodes of the network. Experiments on synthetic and real-world networks using the SIR diffusion model demonstrate the effectiveness the \textit{GLR} strategy in comparison with the Closeness, Degree, Betweenness and k-shell based strategies. Berahmand et al. \cite{dil} proposed the Degree and Clustering coefficient and Location strategy \textit{DCL}. It immunizes the best spreaders based on a combination of the degree and the inverse cluster coefficient of a given node. The latter two measures are also combined with the degree of its neighbors and the common links between the node and its neighbors to define the location of a node (whether it is in the core or the periphery of the community). This strategy allows identifying low-degree bridges and some critical hub nodes. Comparisons based on the SIR and the SI models reveal that the proposed method outperforms the well-known strategies such us the Degree, Betweenness, Eigenvector, PageRank and the k-shell based strategies. The Community Hub-Bridge strategy \cite{ourstr} is based on a linear measure. It is a weighted combination of the number of intra-community links and the number inter-community links. The first term of this measure is weighted by the size of the community. The aim of this is to prioritize the immunization of hubs located in large communities due to their big influence. The second term of the expression is weighted by the number of neighboring communities to target in priority bridges having many connections with multiple communities. According to SIR simulations performed on synthetic and real-world networks, this strategy is more efficient than the Number of Neighboring Communities, Community Inbetweenness, CBM and Comm strategies. It is particularly suited for networks with strong community structure (having a small proportion of inter-community connections). The Weighted Community Hub-Bridge strategy \cite{ourstr} is another variant of the previous strategy. It is based on a linear measure weighted also by the density of the inter-community links.  It is weighted such that, in networks with strong community structure,  more importance is granted to bridges while in networks with loose community structure more importance is given to the local community hubs. Experimental results show that it outperforms the previous strategy namely in networks with loose community structure.

\subsubsection{Modular centrality}
The above-mentioned immunization strategies are based on measures that quantify either the global influence of nodes by selecting bridge nodes, or the local influence of nodes by targeting community hub nodes. Other centrality measures highlight nodes having both local and global influence for immunization. The modular centrality considers two types of influences for a node in a modular network: A local influence on the nodes belonging to its own community through the intra-community links, and a global influence on the nodes of the other communities through the inter-community links. Therefore, in this approach, centrality measures are not represented by a simple scalar value but rather by a two-dimensional vector, the so called \textit{Modular centrality}  \cite{modular}. Its first component measures the local influence of the node, while the second component measures its global influence. The Modular centrality is computed following two steps. The global component of the vector is computed on the global network obtained by removing all the intra-community links from the original network. Remaining isolated nodes are also removed. The local component is computed on the local graph obtained by removing all the inter-community links from the original network.
The Modular centrality is computed according to the following algorithm:

Step 1. Choose a standard centrality measure $\beta$.

Step 2. Remove all the inter-community edges from the original network $G$ to obtain the set of communities $\mathcal{C}$ forming the local network $G_{l}$.

Step 3. Compute the local measure $\beta_{L}$ for each node in its own community.

Step 4. Remove all the intra-community edges from the original network to reveal the set of connected components $\mathcal{S}$ formed by the inter-community links.

Step 5. Form the global network $G_{g}$ based on the union of all the connected components. Isolated nodes are removed from this network and their global centrality value is set to 0.

Step 6. Compute the global measure $\beta_{G}$ of the nodes linking the communities based on each component of the global network.

Step 7. Add $\beta_{L}$ and $\beta_{G}$ to the Modular centrality vector $B_{M}$.

This approach allows to redefine all  the standard centrality measures designed for non-modular networks to networks with non-overlapping community structure. A series of experiments have been performed on both real-world and synthetic networks using the SIR model in order to investigate the efficiency of the Modular centrality. Results show that the Local measure is more efficient in networks with strong community structure, while Global measure performs better in networks with a weak community structure. Furthermore, the measure that combines both components outperforms the local, the global and the classical measure. Recently this work has been extended to networks with non-overlapping community structure \cite{modular2}.


\subsubsection{Discussion}
Comparing with stochastic immunization strategies, the epidemic size of deterministic strategies (e.g., \textit{Comm}, \textit{CbM}, \textit{CBH}, 
\textit{WCBH} and \textit{NNC}) outperforms \textit{CBF} and \textit{BHD} methods in all the networks. Indeed, stochastic strategies only seek current node's information, while deterministic strategies require the access to the whole network structure. That explains why the performance of stochastic strategies is usually far from the deterministic ones.
To compare  the performance of deterministic strategies, we consider two extreme cases:  Networks with well-defined community structure and networks with weak community structure. In  the first case, the communities are very separated from each other. Hence, there are few inter-community connections between the different modules of the network. The local strategies have proven to be more efficient than the global strategies in such networks. The Super node strategy outperforms some global strategies such as the global betweenness method. Actually, there is a great chance that the epidemic stays confined inside the communities because of the small number of inter-community links. Therefore, immunizing hub nodes or community core nodes may appear as the most efficient way to stop the epidemic diffusion in networks with strong community structure.
In networks with medium or unclear community structure,  there are a large amount of inter-community connections in the network. The epidemic in this case can move easily from one community to another. Thus, bridge nodes may play a major role in the diffusion process. That explains the efficiency of the global strategies as compared to the local ones in these networks. The Number of Neighboring Communities (\textit{NNC}) for instance is more efficient than the local degree and the super node strategies. The combination-based strategies, on the other hand, target both type of nodes. They are overall more efficient than both local and global strategies in networks with different structures. Some strategies such as \textit{CbM}, \textit{CBH} and \textit{WCBH} outperform the super node, the local and  the global betweenness and degree-based strategies. Furthermore, \textit{WCBM} has proven to be more efficient than some other combination-based strategies (e.g., \textit{Comm}, \textit{CbM} and \textit{CbC}). This strategy uses different level of information about the topological properties of the community structure such as the size of communities, the number of neighboring communities of each node and the proportion of inter-community links of each community. Thus, it uses more information about the community structure as compared to the other strategies. Therefore, the performance of the immunization strategies increases when more information about the community structure is used. 

These assumptions led to the introduction of the \textit{Modular centrality}, which is a bi-dimensional vector measuring both local and global influence of each node in the network. This approach investigated for some classical centrality measures (Degree, Betweenness, Closeness and Eigenvector) shows that the Local measure is more efficient in networks with strong community structure, while the Global measure performs better in networks with loose community structure. Moreover, the performance of ranking strategies combining both components of the Modular centrality is more efficient than using only one component. Furthermore, better results were even obtained by using more information related to the topological properties of the community structure. These experimental results of the \textit{Modular centrality} confirm the ones obtained with the alternative deterministic strategies.

\subsection{Overlapping community structure}
Communities can often overlap in real-world networks. In this case, nodes can belong to more than one community at once. Identifying such overlapping nodes is crucial for controlling the epidemic spreading. These nodes can extend the epidemic diffusion across all communities to which they belong. Some strategies select these nodes for immunization. Hebert \textit{et al.} \cite{membership} proposed a straightforward strategy which directly counts the membership number of each node in the network. Chakraborty \textit{et al.} \cite{stra3} analyze how immunization based on the membership number of overlapping nodes affect the largest connected component size. OverlapNeighborhood \textit{ON} \cite{manish} is another strategy that targets the neighbors of the overlapping nodes for immunization. It is based on the idea that overlapping nodes are connected to many hub nodes located in the different communities to which they belong. The Membership and OverlapNeighborhood strategies are defined as follows:\\

\noindent\textbf{Membership strategy:} This strategy \cite{membership} is applied to networks with overlapping modular structure. It is based on a measure that counts simply the number of communities to which a node belongs. If the membership of a node $i$ is greater than $1$, i.e., this node belongs to an overlapping region in the network. Experimental results using the SIR model have shown that this strategy outperforms degree, coreness and betweenness-based strategies in networks with denser communities and by using a higher infection rates.\\

\noindent\textbf{OverlapNeighborhood strategy (ON):} This method \cite{manish} selects immediate neighbors of overlapping nodes as the top influential spreaders. Its main objective is to select the most highly connected nodes using a limited amount of information at the community level. Indeed, there is a high probability that nodes with very high connections are neighbors to overlapping nodes since they are part of more than one community. This is also due to the power-law degree distribution in real-world networks.  The simulation results  revealed that this method outperforms CBF, BHD and RWOS methods. It performs better or as good as Degree and Betweenness centrality based methods using less information about the overall network.\\

The Overlapping constraint coefficient (\textit{OC}) \cite{oc} is an immunization strategy that highlights the influential nodes based on the multiplication of two measures. The first measure represents the membership of a given node which quantifies its propagation capacity. So, the more communities a node belongs to, the more communities the node can influence. The second measure represents the network constraint coefficient of the node, which quantifies its propagation speed in the communities. SIR simulations demonstrate that the Overlapping constraint coefficient strategy outperforms the Degree, Betweenness, Closeness and the k-shell based strategies. The Influence Maximization based on Label Propagation Algorithm (\textit{IM-LPA}) \cite{impla} is another strategy designed for networks with overlapping communities. It is based on an improved version of the Label propagation algorithm \cite{lpa}. It operates in two phases: the seeding phase and the label propagation phase. At the beginning of the seeding phase, the set of seed nodes is empty and all the nodes of the network are considered as candidate nodes. After that, the node with the highest degree is added to the seed set and all its neighbors are removed from the candidate node set. This process is repeated until the candidate node set becomes empty. This phase guarantees that the selected seed nodes are independent from each other. In the label propagation phase, each seed node is associated with a unique label. Then, the labels expand from the seed nodes until covering all the other nodes of the network. Nodes may have several labels. Thus, they can belong to several communities. At the end of this process, the centrality of each node can be measured by the number of nodes associated to its label. Nodes with the highest measure can propagate the epidemics to a large set of nodes of their communities. The Independent cascade diffusion model (\textit{IC}) was performed on both synthetic and real-world networks. Results demonstrate the efficiency of the \textit{IM-LPA} strategy in identifying the influential spreaders as compared to the Degree, Betweenness, Closeness, K-shell and PageRank-based strategies.

\section{Summary and Conclusions}

In complex networks, community structures are widely observed. Despite the fact that this property is well-recognized, it is very often ignored when it comes to use it in order to develop new techniques in the field. In this paper, we consider three hot topics linked to the community structure of complex networks. First one focuses on the fundamental issue of community detection in static networks.
The second one discusses the same issue but for temporal networks. Finally, the third one examines immunization strategies designed for modular networks.  

After the introduction, the second section focuses on static networks in which detecting communities can be viewed as partitioning of the network into clusters in which the nodes are more densely connected to each other than to the nodes in the rest of the network.  In this section, we look at community detection based on this fundamental assumption about community structure. 

In summary, the current state of the art in this area is as follows. One systematic approach to community detection is to select a metric of community quality and maximize it. Several of such metrics~\cite{Chen2014Community, Lu2018Adaptive,lewis2010function,simon1991architecture,porter2009communities,reichardt2006statistical} are variants or improvements based on the modularity metric of community structure that measures the difference between the observed fraction of edges within a community and this fraction expected in a random graph with the same number of nodes and the same degree sequence. That gave raise to modularity maximization~\cite{newman2006Modularity} as one of the state-of-the-art methods for community detection. However, it suffers from the so-called resolution limit problem~\cite{fortunato2007resolution,lancichinetti2011limits}, a tendency of standard modularity to increase when some small well-formed communities are combined into inappropriate large clusters, while some large well-formed communities are spread among smaller ones. Some of the above mentioned variants of the modularity function have been proposed to either resolve this problem~\cite{Chen2014Extension,Lu2018Adaptive} or to enable detection of communities at different scales~\cite{lewis2010function, simon1991architecture, porter2009communities}. A popular choice for the latter is the {\em generalized modularity} of Reichardt and Bornholdt~\cite{reichardt2006statistical}, which scales the discovered community sizes according to a simple resolution parameter. This parameter is not fixed in the definition of the generalized modularity. Hence, many approaches~\cite{fenn2009communities,mucha2010communities,Traag2013Significant} try different values of the resolution parameter to find proper community structures in the real networks. When the resolution parameter is set as one, the generalized modularity reduces to the traditional modularity.  Another drawback of this approach is that the stochastic block model requires the selection of the number of communities, because selecting a large number of blocks always leads to a high likelihood of generating the observed network. Another drawback of this approach is that the stochastic block model requires the selection of the number of communities, because selecting a large number of blocks always leads to a high likelihood of generating the observed network. Therefore, recent works~\cite{riolo2017efficient, newman2016estimating,peixoto2017bayesian} adopt Bayes model selection to find the appropriate number of communities in a network. According to Occam's Razor, this approach also minimizes the description length (MDL) of the block model~\cite{peixoto2012entropy, peixoto2017bayesian} so that community detection algorithm finds the most suitable number of communities.

An extension of this model~\cite{karrer2011stochastic} introduces the so-called degree-corrected stochastic block model in which the node degrees are also used as parameters, making the expected node degree in the model equivalent to the observed node degree. Since the nodes in the same community tend to have broad degree distributions, this simple yet effective extension of node degrees improves the performance of the models for statistical inference of community structure in the real-world networks. The degree-corrected planted partition model is a special case of the degree-corrected stochastic block model.

Recently, Newman~\cite{newman2016equivalence} proved partial equivalence of the two approaches by showing that modularity maximization is equivalent to the maximum-likelihood estimation (MLE) of the degree-corrected planted partition model on the same graph. Lu and Szymanski~\cite{lu2019asymptotic} established an asymptotic theoretical upper and lower bounds on the resolution parameter of generalized modularity. When the upper bound larger than the lower one then we know that there is a resolution parameter that avoids modularity resolution problem in the corresponding network.  The open question now is how to proceed if the upper bound is smaller than the lower one. 

An alternative approach to metric maximization is the statistical inference that fit the generative model to the observed network data. Such approach assumes the observed network is produced by a random graph model with a pre-defined partition of the network as the model parameter. In general, the statistical inference aims at recovering the partition which maximizes the likelihood of the random graph model generating the observed network data. One widely used generative model for community structure is the stochastic block model~\cite{karrer2011stochastic} where nodes are organized as blocks and edges are placed between the nodes independently at random, with a probability depending on the block assignments of the endpoints. The weakness of this approach is that the model considers nodes in the same block statistically indistinguishable from each other, so the most likely block assignment often groups the nodes of similar degrees in a block, resulting in lower and higher-degree blocks, rather than the traditional community structures. Moreover, the inference is actually much more complicated than maximizing generalized modularity. One of the reasons is that the current versions of stochastic block model searches through the large space of potential solutions containing both assortative and disassortative structures~\cite{Peele2017ground}. Consequently, inference algorithms using these models are often trapped in a solution unsuitable for the user and it takes them long time to escape. To address this issue, the authors of \cite{lu2019Regularized} apply a simple constraint on nodes internal degree ratio in the objective function.

Despite the significant progress made towards community detection using fundamental properties of the communities, provably optimal algorithms are still beyond our reach for the modularity maximization based approaches. The current open question is how to proceed if for the network in question no single resolution parameter exists that will allow modularity maximization to avoid anomalies. At least there is now a simple test, introduced in \cite{lu2019asymptotic}, that allows for detecting such cases and proposes a method for finding  a solution free of such anomalies.

In the third section of the paper, we briefly overview the most popular approaches and recent advances in the field of evolving community detection. Nowadays, the availability of time stamped or time dependent data on networked systems is becoming widespread, hence the scientific interest towards the study of time evolving networks is increasing. Locating communities in time dependent networks is a non-trivial and challenging problem, with an impressive number of proposed different solutions. 

A relatively straightforward idea is to represent the time evolving network as a sequence of static snap-shots, and apply one of the well-known static community detection algorithms on the series of static graphs, as was done in Refs.\cite{Hopcroft_evolv_coms,Asur_evolv_coms,Palla_com_evolv,Greene2010TrackingTE,Martin_evolv_coms,Brodka_GED}. Naturally, the obtained communities have to be matched at subsequent time steps in order to obtain time evolving clusters. The advantage of this approach is that basically any static community finding method can be used, however the drawback is that the matching part can become complicated and the threads of the evolving communities may turn needlessly intricate. 

In contrast to snap-shot based methods, the concept of evolutionary algorithms treats the inference of the time dependent communities in a unified framework. Indeed, in this case, the structure of a community at a given time step $t$ can be influenced by information coming from other time steps as well \cite{Chakrabarti_evolutionary_clustering,Chi_evolv_spectral_clust,Lin_evolv_coms,Zhou_evolv_coms,Tang_evolv_coms,Folino_evolv_coms,Sun_evolv_coms,Gong_evolv_coms,Kawadia_evolv_coms,Crane_evolv_coms,Gorke_dyn_coms_smooth}. A popular approach along this line is to formulate the aim for a smooth evolution over time together with the goal of obtaining precise communities reflecting the true modular structure of the network at any time point as an optimization problem. Further methods following a similar track are based on multislice networks \cite{Mucha_multiplex_coms}, consensus clustering \cite{Lancichinetti_consensus_clust}, or generative models such as the stochastic block model \cite{Yang_dynamic_SBM,Tiago_time_varying,Peel_detectability,Tiago_and_Martin_dynamic_coms,Leto_unobserved_detection,Tiago_dynamic_reconstruction}. 

A closely related idea to the above is given by incremental clustering \cite{Aynaud_increment_clust,Ning_increment_clust,Bansal_increment_coms,Gorke_increment_clust,Xie_and_Bolek_increment_clust,Cazabet_increment_clust,Duan_increment_clust,Falkowski_increment_clust,Nguyen_overlap_increment_clust,Cazabet_and_Amblard_increment_clust,Gorke_incr_clust_2,Ma_increment_clust,Lee_increment_clust,Zakrzewska_increment_coms}, where only the time steps relatively in the past are taken into account when extracting the communities at a given date $t$ . Although this is somewhat a more restrictive setup compared to evolutionary clustering, the advantage of this approach is that it enables in principle the online clustering of networks \cite{Aggarwal_online_clustering,Zanghi_online_clustering,Rossetti_online_coms,Tan_online_coms}. Besides online community detection, the concept of forecasting the future events and changes in time dependent communities is also gaining considerable interest \cite{Leskovec_predict,Patil_predict,Goldberg_predict,Brodka_predict,Gliwa_predict,Takaffoli_predict,Saganowski_predict}. 

Partly due to the large number of different methods, providing a well-controlled benchmark system on which the proposed algorithms can be tested and compared has become a very important challenge as well. However, this problem is relevant also from other perspectives, such as e.g., measuring the quality of the obtained evolving communities. A number of important first steps have already been made in this direction, such as the introduction of the time dependent version of the static Girvan-Newman benchmark \cite{Girvan_Newman} in Ref.\cite{Lin_evolv_coms}, the dynamic modification of the static LFR benchmark \cite{lfr} in Ref.\cite{Greene2010TrackingTE}, and the proposition of a benchmark based on a time evolving stochastic block model in Ref. \cite{Granell_benchmark}. Furthermore, the problem can be also brought into a more general context with the concept of multilayer community benchmarks \cite{Bazzi_benchmark}, while tailor made benchmarks specific for a given problem or method can be also well motivated \cite{Rossetti_benchmark}. 

Nevertheless, how to measure and compare the performance of evolutionary community finding algorithms is a highly non-trivial question, related to which further advances can be expected in the future. What makes the problem especially difficult is the rather diverse nature of both the time evolving networks and of the applied methods. There are systems where we find quite large variations in the network structure across subsequent time steps, whereas other networks show a gradual, significantly smoother evolution in time; and in respect of the proposed algorithms, there are methods concentrating more on the accuracy of the obtained communities, whereas others focus instead on the smoothness and coherence of the evolution. Based on that, defining e.g., a quality function analogous to modularity is far from trivial, and bringing the field to a common ground in terms of benchmarks and comparison provide interesting and important challenges for the future.

In the fourth section, we look at how the community structure affects the diffusion process of epidemics, and how to use information about the community structure in order to design effective  immunization strategies to control epidemics in modular networks. We can distinguish two main approaches to solve this issue. The first is the stochastic approach beneficial when little is known about the full network structure or when the networks are too large to compute features for each nodes. However in general, the second approach of using the deterministic strategies outperforms the non-deterministic strategies.

Overall, the works presented above demonstrate that it is important to consider the community structure of real-world networks to develop more suitable immunization strategies.  Some stochastic strategies are designed to target the nodes linking the communities (bridges) because such nodes connect to many parts of the network. Others concentrate on the highly connected nodes (hub). A third type of strategy targets bridges and hubs. Globally their effectiveness depends of the community structure strength. Indeed,  the best strategies are the ones that give more importance to the hubs when there is a small proportion of links between the communities. But when the proportion of inter community links increases, it is better to immunize the bridge nodes first. So, there is a need for  new stochastic strategies that can adapt to both situations and can be tuned according to the community structure strength.
In fact, the performance of stochastic strategies increases when additional knowledge about the community structure of the network is utilized by the algorithm. 
Globally, deterministic strategies are more sophisticated than stochastic strategies.
since they can easily exploit knowledge about the network topology. We classified them into three categories. Local strategies concentrate on the information into the communities, while global strategies use the information between the communities. Finally global and local strategies exploit both type of knowledge. We observe the same behavior that the one observed with stochastic strategies. Indeed, local strategies outperform the global strategies in networks with well-defined community structure while global strategies are more effective in networks with loose community structure. Strategies exploiting both aspects perform generally better. Indeed, they incorporate in their definition additional information about the community structure as compared to local or global strategies.
Finally, we believe that the modular centrality framework is very promising. It gives a clear idea of how to use both local and global knowledge of the community structure. Additionally, as there is no constraint about the centrality used and the way to combine both dimensions, there is room for improvement.

In networks with overlapping communities, immunization strategies take also into account the overlapping nodes which belong to multiple communities. These strategies show the importance of these nodes, and show also their ability in terms of the spread of infections. 
 The \textit{OC} strategy has proven to be the most effective deterministic strategy based on overlapping nodes. Indeed, this strategy considers other information about the community structure as compared to the membership, OverlapNeighborhood and the \textit{IM-LPA} strategies. It is a combination-based method. It targets nodes having  access to multiple communities and with high propagation speed in these communities.  The stochastic strategy RWOS compares well with its alternatives. However we cannot call it a pure stochastic strategy, because the overlapping nodes need to be known or estimated.

All of these works give us a sense of directions for designing new immunization strategies tailored to the network topology. The community structure cannot be ignored and much more knowledge about the formation of the communities and of their main features \cite{gunce4} need to be uncovered and integrated into the immunization strategies in order to better identify the influential nodes. One of the main challenge is to initiate  research concerning semi stochastic strategies such as  RWOS. Indeed, stochastic strategies are the ones that are the more suitable when the network is partially unknown, or too large to uncover its community structure. However, adding information about the community structure make them more effective. That is why the main stream of improvement is in between the effectiveness of the deterministic strategies and the computational efficiency of the stochastic strategies. 

\section*{Declarations}

{\bf Availability of data and material}

\noindent
All data used in this article is publicly available at the websites cited in the references. No program source code is described in
the paper.
\\

\noindent
{\bf Competing interests}

\noindent
The authors declare that they have no competing interests.
\\

\noindent
{\bf Funding}

\noindent
GP was partially supported by the European Union’s Horizon 2020 Research and Innovation Programme under Grant Agreement No. 740688 and by the National Research, Development and Innovation Office under Grant No. K128780. BKS was partially supported by the Army Research Laboratory under Cooperative Agreement No. W911NF-09-2-0053 (the Network Science CTA), and the Office of Naval Research (ONR) Grant No. N00014-15-1-2640.
\\

\noindent
{\bf Acknowledgment}

\noindent
The authors wish to acknowledge a partial support from the European Union’s Horizon 2020 Research and Innovation Programme under Grant Agreement No. 740688, from the Hungarian National Research, Development and Innovation Office under Grant No. K128780, from the U.S. Army Research Laboratory under Cooperative Agreement No. W911NF-09-2-0053 (the Network Science CTA), and from the U.S. Office of Naval Research (ONR) Grant Mo. N00014-15-1-2640.
\\

\noindent
\subsection*
{\bf Authors' contributions}

\noindent
All authors conceived of the ideas of the study. BKS and XL prepared and wrote the section 2, titled ``The random graph models for community detection''. GP prepared and wrote the section 3 titled: ``Time evolving communities''. HC prepared and wrote the section 4 titled: ``Immunization strategies''. All authors prepared and wrote section 1, titled ``Introduction,'' and section 5, ``Summary and Conclusions''. All authors read, edited and approved the final manuscript.\\

\noindent
{\bf Authors' information}\\

\noindent
{\bf Hocine Cherifi} is a Professor of Computer Science at the University of Burgundy, Dijon, France.

\noindent
{\bf Gergely Palla} is a Senior Research Associate in the Statistical and Biological Physics 
Research Group of Hungarian Academy of Science at the Eotvos University, Budapest, Hungary. 

\noindent
{\bf Boleslaw K. Szymanski} is the Director of Network Science and Technology Center, the Claire and Roland Schmitt Distinguished Professor of Computer Science, and a Professor of Physics at the Rensselaer Polytechnic Institute.

\noindent
{\bf Xiaoyan Lu} is the fourth year graduate student at the Network Science and Technology
Center and the Department of Computer Science at the Rensselaer Polytechnic Institute.


\bibliographystyle{spphys}       
\bibliography{pos_pap}   

%
%

\end{document}